\documentstyle[12pt]{article}
\begin{document}
\baselineskip = 21 pt
\thispagestyle{empty}
\title{
\vspace{-1.5cm}
\begin{flushright}
\begin{tabular}{c c}
 & { \normalsize MPI-Ph/93-35}\\
 & {\normalsize July 1993}
\end{tabular}
\end{flushright}
\vspace{1.cm}
Unification of Couplings and the Top Quark Sector
in the Minimal Supersymmetric Standard Model
\thanks{ To appear in the Moriond Proceedings on
$\lq \lq $ Electroweak Interactions and Unified Theories", Les Arcs,
France, March 1993.}
{}~\\}
\author{{\bf
 Marcela Carena}   \\
{}~\\
Max-Planck Institut f\"{u}r Physik -
Werner-Heisenberg Institut -\\
D-80805 M\"{u}nchen, Fed. Rep. of Germany.\\
{}~\\
{}~\\
{}~\\
{}~\\ }
\date{
\begin{abstract}
I analyze the
predictions for the strong gauge coupling,
$\alpha_3(M_Z) $, and
the top quark and light
Higgs masses in the
framework of gauge and bottom-tau Yukawa coupling unification in the
minimal supersymmetric standard model.
 These
predictions depend on the effective supersymmetric threshold scale
 $T_{SUSY}$, which is only very slightly dependent on the squark and
 slepton masses, and strongly dependent on the Higgsino masses as well
as on the mass ratio of the gauginos of the strong and weak interactions.
Within the minimal supersymmetry
 breaking scheme and for supersymmetric masses below or of the order of
 1  TeV, I obtain $\alpha_3(M_Z) \geq 0.116$, while, if the
running bottom quark mass at the physical mass
 is constrained to be $ m_b(M_b) \leq 4.1 $GeV,
then, for moderate values of $\tan \beta$,
perturbative unification is achieved only if
 $\alpha_3(M_Z) \leq 0.124$. Unification of gauge and
 bottom-tau Yukawa couplings   yield
predictions for the top quark mass, 140 GeV $\leq M_t \leq 210 $ GeV
  for $1 \leq \tan \beta \leq 30$, which are
remarkably close to the
infrared quasi fixed point  values for this quantity. For
 the light Higgs mass I obtain
 $m_h \leq 130 (165)$ GeV
if the characteristic squark mass is below 1 (10) TeV.
\end{abstract}}
\maketitle
\newpage
Recent studies have shown that, within the minimal supersymmetric
extension of grand unified theories,
  it is possible to extrapolate
 the standard model gauge couplings to obtain unification
 at some high energy
scale $M_{GUT}$ in a framework   compatible with
present experimental data$^{1)-4)}$.
Furthermore, the
 unification of  couplings   in the Yukawa sector appears
naturally in some grand unified  models, yielding definite
predictions   for the top quark mass$^{5)-7)}$.
For the extent of this talk I
shall concentrate on the Minimal Supersymmetric Standard Model (MSSM)
within the minimal presumtion of grand unification,
 without any specific
assumptions
about the physics above the unification scale.
However, in order
to consider the problem of unification of couplings in a proper way
it is necessary to analyze   the incidence   of possible threshold
corrections at $M_{GUT}$, which
depend
on the specific grand unified scenario at the high energy
scales. Thus,
I shall  also briefly
discuss the relaxation of the unification condition for the
couplings and its relation to the infrared quasi fixed
 point predictions
for the top quark mass.

In a  first approximation to the
problem of unification of gauge couplings,
it is usual to assume that all the  supersymmetric particles have a
common mass $M_{SUSY}$. In such extreme case, the common mass scale
 $M_{SUSY}$ coincides with the scale $T_{SUSY}$  which,
  in general, denotes the scale characterizing the supersymmetric
threshold corrections to the gauge couplings. Within this context, for
values of the low energy parameters consistent with experimental data,
the unification of gauge couplings may be achieved for a unification
scale $M_{GUT} = {\cal{O}} (10^{16} GeV)$  and a common sparticle mass
$M_{SUSY}$
of the order of the weak scale$^{1)}$.
However, once the condition of equality
of all the supersymmetric
particle masses is relaxed, the actual scale characterizing the
supersymmetric threshold corrections to the gauge couplings may
significantly differ from any of the sparticle masses$^{2)-4), \;7)}$.
 Nevertheless, as I shall show below,
the consequences of an arbitrary supersymmetric spectrum on the
low energy predictions for the gauge couplings can always be parametrized
in terms of a single threshold scale $T_{SUSY}$ $^{4), \; 7)}$.

Considering a general supersymmetric spectrum, each sparticle $\eta$
contributes to the one loop supersymmetric threshold corrections
to $1/\alpha_i(M_Z)$ with a factor $b_i^{\eta}$ yielding,
\begin{equation}
\frac{1}{\alpha_i^{Sthr}(M_Z)} =
\sum_{\eta, M_{\eta} > M_Z} \frac{b_i^{\eta}}{2 \pi} \ln\left(\frac{
M_{\eta}}{M_Z} \right).
\label{eq:Sthr}
\end{equation}
Assuming the unification of the three gauge couplings at $M_{GUT}$
and considering their two loop beta functions,
 one coupling, for example $\alpha_3(M_Z)$, is determined
 as a function of the electroweak gauge couplings and the above
supersymmetric threshold corrections.
\begin{equation}
\frac{1}{\alpha_3(M_Z)} = \frac{1}{\alpha_3^{SUSY}(M_Z)}
+ \Delta^{Sthr} \left(\frac{1}{\alpha_3(M_Z)}\right) ,
\label{eq:alpha3}
\end{equation}
where
\begin{equation}
\frac{1}{\alpha_3^{SUSY}(M_Z)} = \frac{b_1-b_3}{b_1-b_2} \left[
\frac{1}{\alpha_2(M_Z)}
+ \gamma_2 + \Delta_2\right] -
 \frac{b_2-b_3}{b_1-b_2} \left[
 \frac{1}{\alpha_1(M_Z)} + \gamma_1 +
\Delta_1\right] -
  \gamma_3 - \Delta_3
\end{equation}
gives the value of the strong gauge coupling at $M_Z$ if the
theory were supersymmetric all the way down to $M_Z$, with $b_i$
the one loop beta function coefficient  of the gauge coupling $\alpha_i$
in the MSSM, $\gamma_i$ the two loop contributions to the beta functions
and $\Delta_i$ a constant term associated to scheme conversion.
The last term in eq. (\ref{eq:alpha3}) accounts for the supersymmetric
threshold corrections$^{4),7)}$
\begin{equation}
 \Delta^{Sthr}\left(\frac{1}{\alpha_3(M_Z)}\right)   =
 \frac{(b_1-b_3)}{(b_1-b_2)} \frac{1}{\alpha_2^{Sthr}} -
 \frac{(b_2-b_3)}{(b_1-b_2)} \frac{1}{\alpha_1^{Sthr}} -
 \frac{1}{\alpha_3^{Sthr}} =
      \frac{19}{28 \pi} \ln \left(\frac{T_{SUSY}}{M_Z} \right),
\label{eq:DSthr}
\end{equation}
which can be parametrized in terms of the
supersymmetric effective threshold scale $T_{SUSY}$. Therefore,
it is most useful to analyze
the exact dependence of $T_{SUSY} $ on the physical masses of the
supersymmetric particles.

Considering different characteristic  masses for  the  squarks,
 $m_{\tilde{q}}$, gluinos,
 $m_{\tilde{g}}$, sleptons,
 $m_{\tilde{l}}$, electroweak gauginos,
 $m_{\tilde{W}}$, Higgsinos,
 $m_{\tilde{H}}$, and the heavy Higgs doublet,
 $m_H$, the following expression
 is derived$^{7)}$,
\begin{equation}
T_{SUSY} =
 m_{\tilde{H}}
  \left( \frac{
 m_{\tilde{W}}}{m_{\tilde{g}}} \right)^{28/19}
\left[
  \left( \frac{
 m_{\tilde{l}}}{m_{\tilde{q}}} \right)^{3/19}
  \left( \frac{m_H}{m_{\tilde{H}}} \right)^{3/19}
  \left( \frac{
 m_{\tilde{W}}}{m_{\tilde{H}}} \right)^{4/19}
\right] .
\label{eq:TSUSY}
\end{equation}
The above expression is very useful\footnote{This expression holds
whenever all the particles have
masses above $M_Z$. If, instead, any of the  masses is
below $M_Z$,
it should
be replaced by $M_Z$ for the purpose of computing $T_{SUSY}$.}. It shows
that $T_{SUSY}$
has only a slight dependence on the squark and slepton masses as well
as on the heavy Higgs mass, but, instead, a very strong dependence
on the  Higgsino and  gaugino masses.
In models in which the source of supersymmetry breakdown is given
by a common gaugino mass $M_{1/2}$ at the unification scale and assuming
that there is no significant gaugino-Higgsino mixing - the
overall Higgsino mass scale is approximately given by the absolute value
of the renormalized supersymmetric mass parameter $\mu$ - $T_{SUSY}$,
 eq. (\ref{eq:TSUSY}), can
be approximately given by,
\begin{equation}
T_{SUSY} \simeq
 m_{\tilde{H}}
  \left( \frac{
 \alpha_2(M_Z)}{\alpha_3(M_Z)} \right)^{3/2} \simeq
 \frac{|\mu|}{7}.
\label{eq:TSUSYmu}
\end{equation}
Therefore, if $|\mu|\leq 1  $ TeV it follows $T_{SUSY} \leq {\cal{O}}
(M_Z)$.

In order to perform a self consistent two loop
Renormalization Group (RG)
analysis it is most important
to consider the experimental data in such a way to determine the low
energy boundary conditions as precise as possible.
The experimental prediction for $\sin^2\theta_W $ in the modified
$\bar{MS}$ scheme depends quadratically on the top quark mass$^{4)}$,
\begin{equation}
\sin^2 \theta_W(M_Z) \simeq 0.2324 -   10^{-7}  GeV^{-2}
 (M_t^2 - (138 GeV)^2)\; .
\label{eq:sintheta}
\end{equation}
However, it is also possible  to consider the best fit to the data with
a free
top quark mass value, $\sin^2\theta_W (M_Z)= 0.2324 \pm 0.0012$.
The electromagnetic gauge coupling,
 $1/\alpha(M_Z) = 127.9$, has instead only a
logarithmic dependence on the top mass.
 Contrary to the situation in the
electroweak sector, the    strong gauge coupling
is not so precisely known. A conservative attitude is to take
$\alpha_3(M_Z) = 0.12 \pm 0.01$. The tau mass is taken to be $M_{\tau}
= 1.78 $ GeV. Concerning the bottom quark sector, I consider the range
of values for the physical bottom quark mass quoted in the particle data
book, $M_b = 4.7 - 5.2 $ GeV$^{8)}$.
 Observe that, after including
QCD corrections at the  two loop level, the running mass differs
significantly from the physical mass. In fact, the above range for
$M_b$ corresponds to a running mass
 $m_b(M_b) \simeq  4.1 - 4.6 $ GeV. Similarly, the physical top
quark mass $M_t$ is approximately 6 $\%$ larger than the corresponding
 running mass $m_t(M_t)$.

In the framework of gauge coupling unification, neglecting threshold
corrections at $M_{GUT}$, I obtain that the value of the strong gauge
coupling increases for smaller values of $\sin^2 \theta_W$ as well
as for smaller values of $T_{SUSY}$.
At the two loop level, the strong gauge coupling receives
negative contributions from the top quark Yukawa coupling, $h_t$. $\;$
In  table 1 $^{7)}$, $\;$I show the above behaviour
for values of $Y_t = h_t^2/4 \pi \simeq 0.4 - 1$.
Thus, within the minimal supersymmetry
 breaking scheme and for supersymmetric particle masses  below
 or of the order of 1 TeV, in  which case the value of $T_{SUSY}$,
 eq. (\ref{eq:TSUSYmu}), is constrained to be $T_{SUSY} \leq 200$ GeV,
  a lower bound on
the strong gauge coupling, $\alpha_3(M_Z) \geq 0.116$, is derived.
{}~\\
\begin{center}
\baselineskip = 17 pt
Table 1. Dependence of $\alpha_3(M_Z)$ on $\sin^2 \theta_W(M_Z)$
and  $T_{SUSY}$
in the framework of two loop RG analysis with gauge and
Yukawa coupling unification,
for $m_b(M_b) = 4.3$GeV.\\
{}~\\
\begin{tabular}{|c|c|c|}
\hline \
$\sin^2 \theta_W(M_Z)$
&$\alpha_3(M_Z)$ for $T_{SUSY} = 1$ TeV
 &$\alpha_3(M_Z)$ for $T_{SUSY} = 100$ GeV
\\  \hline
$0.2335
$
 &0.111  &0.118
\\ \hline
$0.2324$
 &0.115  &0.122
\\ \hline
$0.2315$
 &0.118  &0.126
\\ \hline
\end{tabular}
\\
{}~\\
\end{center}
\baselineskip = 21 pt

The inclusion of the unification condition    for the
 bottom  and tau Yukawa couplings leads to an interesting result:
The top quark mass values computed for values of the bottom quark mass
compatible with present experimental data, $M_b \leq 5.2 GeV$, are
remarkably close to the infrared  quasi fixed point predictions
 associated with
a large  top quark Yukawa coupling,
$Y_t \simeq 0.1 - 1$,  at $M_{GUT}$.
Table 2$^{7)}$ illustrates this property for an intermediate
value of the running bottom quark mass and various values
of $\tan \beta$, the ratio of the vacuum expectation values of  the two
Higgs fields. In fact, the above is the
{}~\\
\begin{center}
\baselineskip = 17 pt
Table 2. Comparison of the two loop RG analysis
 with the approximate infrared fixed point predictions,
for $\sin^2 \theta_W(M_Z) = 0.2324$,
$\alpha_3(M_Z) \simeq 0.122$ and  $m_b(M_b) = 4.3$ GeV.\\
{}~\\
\begin{tabular}{|c|c|c|}
\hline \
$\;\;\;\;\;\;\;\;\;\;\;\;\;$
&IR Fixed Point &RG solution
\\  \hline
$m_t(\tan\beta=1)
$ [GeV]
 &144  &144
\\ \hline
$m_t(\tan\beta=3)
$ [GeV]
 &192  &187
\\ \hline
$m_t(\tan\beta=10)$ [GeV]
 &201  &196
\\ \hline
\end{tabular}
\\
{}~\\
\end{center}
\baselineskip = 21 pt
reason why the predictions
for $M_t$ coming from bottom-tau unification  are basically the same
as those obtained in the framework of supersymmetric top
condensate models, which are also associated with a large $Y_t$  at
the compositeness scale $\Lambda$, if $\Lambda \simeq M_{GUT}$$^{9)}$.
For lower values of the bottom quark mass, the top quark Yukawa coupling
may become too large to be compatible with perturbative unification.
Thus, in order to secure the perturbative consistency of the analysis,
I require that $Y_t(M_{GUT}) \leq 1$ - the two loop corrections are
less than 30 $\%$ of the one loop contributions.
Most generally,
 I obtain that, for values of $m_b(M_b) \leq 4.1 GeV$,
gauge and bottom-tau Yukawa coupling unification requires $\alpha_3(M_Z)
\leq 0.124$ in order to remain in the perturbative region of the top
quark Yukawa sector$^{7)}$.

In Fig. 1, I present the values of the running top quark mass
 as a function of $\tan\beta$ for different values of $\alpha_3(M_Z)$.
It is important to stress that, for any intermediate value
of $\tan \beta = 10 - 30$, a large variation in
$\alpha_3(M_Z) = 0.115 \rightarrow 0.13$ yields a large
 variation
in $\;Y_t(M_{GUT}) \;= \;0.28 \rightarrow \;1$, $\;$which,$\;$ however,
$\;$due to the presence of  the infrared quasi
\vspace{13.0cm}
\begin{center}
\baselineskip = 17 pt
Fig. 1. The running top quark mass as a function of $\tan\beta$
for $m_b(M_b) = 4.3$ GeV, $\sin^2\theta_W(M_Z) = 0.2324$ and
varying the value of the strong gauge coupling  to be
$\alpha_3(M_Z) = 0.115$ (solid line), 0.122 (dashed line) and
0.13 (dot-dashed line). \\
{}~\\
\end{center}
\baselineskip = 21 pt
 fixed
 point, induces only
a very mild increase of about a 5 $\%$ in $m_t$.
Similarly, a mild decrease in $m_t$ follows for larger values of the
bottom quark mass.
Quite generally, I obtain that,
 within the framework of perturbative
gauge and  bottom-tau unification, the  running
top quark mass is bounded to be
below 200 GeV, while values of $m_t = {\cal{O}} (140 GeV)$ may only be
naturally accommodated for low values of $\tan \beta \simeq 1^{7)}$.
In addition, the unification of the three Yukawa couplings of the third
generation at $M_{GUT}$
determines a point in any of the above curves, which  is achieved
for large values
of $\tan \beta \simeq 60$ implying $m_t(M_t) = 165 - 185 $ GeV.
The effect of high energy scale
 threshold corrections to the gauge coupling
unification may change the prediction for $\alpha_3(M_Z)$ in up to
5  $\%^{4)}$.
 However, as shown above, for small and moderate values
of $\tan \beta$,  if the strong gauge coupling remains in the
  range $\alpha_3(M_Z) \geq 0.115$,
  this only induces a slight
variation in the top quark mass values. Concerning the effects of
threshold corrections
 at $M_{GUT}$ to the Yukawa coupling unification condition,
it occurs that a relaxation of exact unification up to
$10 \%$ is possible without destroying the attraction to the
infrared quasi fixed point and, therefore, the stability of the top
quark mass predictions$^{6), 10)}$.

Concerning the Higgs sector, the dominant radiative corrections to the
tree level value of the  lightest CP even
Higgs mass are not only dependent
on  the top quark Yukawa coupling, but on the logarithm of the squark
masses as well $^{11), 12)}$,
\begin{equation}
m_h^2(m_t) \leq  M_Z^2 \cos^2 (2 \beta) + \frac{3}{2 \pi^2} v^2
\sin^4 \beta \; h_t^4(m_t) \;\ln
\left(\frac{m_{\tilde{q}}}{m_t} \right).
\end{equation}
The above expression is for the case in which the CP odd scalar mass
$m_A \gg M_Z$ and    gives the upper values for $m_h$ as a function of
$\tan \beta$.
Since the unification condition only fixes the value of $T_{SUSY}$, which
is only weakly dependent on the squark masses, then, even for fixed
$\tan\beta$ there is no direct determination of $m_h$ within this scheme.
For a given value of $T_{SUSY}$, an upper bound on the
lightest Higgs mass is obtained for  the maximun allowed
value for  $m_{\tilde{q}}$, quite generally, of the order of a few
TeV. In general, I obtain $m_h \leq 130 (165)$ GeV for
 $m_{\tilde{q}} \leq 1 (10)$ TeV $^{7)}$.
Moreover, if
 $m_{\tilde{q}} \leq 1$ TeV, then the experimental bounds on $m_h$
 and $M_t$ almost close the $\tan\beta < 1$ window.

In conclusion, within gauge and bottom-tau Yukawa coupling unification
I obtain
predictions for the strong gauge coupling and $M_t(\tan \beta)$.
 Such
predictions depend on the effective supersymmetric threshold scale
and, therefore, they have a
 strong dependence on the Higgsino and
 gaugino masses. In the above,
I have always considered $\sin^2 \theta_W$ taking
into account the best fit to the data with a free top quark mass.
However, it is important to notice that
considering the correlation between $\sin^2\theta_W $
and $M_t$, eq. (\ref{eq:sintheta}), stringent predictions for
 $\alpha_3(M_Z)$ are obtained. In particular, in the minimal
  supersymmetry breaking scheme, for sparticle masses below or of
  the order of 1  TeV,
the lower bound on the strong gauge coupling becomes
$\alpha_3(M_Z) \geq 0.12$.  Gauge and
 bottom-tau Yukawa coupling  unification
  implies  two loop RG
predictions for the top quark mass,
  145 GeV $\leq M_t \leq 210 $ GeV
  for $1 \leq \tan \beta \leq 30$, which are
remarkably close to its
infrared quasi fixed point  values.
Observe that, taking into account the present  bounds on
the top quark mass value coming from precision measurements,
$M_t \leq 170$ GeV at the one $\sigma$ level, stringent bounds on
 $\tan \beta$  are imposed. This implies that the light Higgs
mass is not allowed to reach its maximun value but is constrained to be
$m_h \leq 80 (90)$ GeV if
 $m_{\tilde{q}} \leq 1 (10)$ TeV. Most important, this implies that
if this scenario with heavy supersymmetric particles holds,
the Higgs will be detected at LEP 200 $^{6), 10)}$.
However, in the case that light supersymmetric particles were present,
a reanalysis of the top quark mass predictions coming from
precision measurements would be necessary. This might loosen the
above bounds on $M_t$  and, therefore, those on
$m_h$ as well.

\end{document}